\documentclass[apl,twocolumn,showpacs,preprintnumbers,amsmath,amssymb]{revtex4}

\usepackage{graphicx}

\begin{document}

\title{Magnetic states of linear defects in graphene monolayers: effects of strain and interaction}

\author{Simone S. Alexandre}
\affiliation{Departamento de F\'{\i}sica, Universidade Federal de
Minas Gerais, CP 702, 30123-970, Belo Horizonte, MG, Brazil}

\author{R. W. Nunes}
\email{rwnunes@fisica.ufmg.br}
\affiliation{Departamento de F\'{\i}sica,  Universidade Federal de Minas
Gerais, CP 702, 30123-970, Belo Horizonte, MG, Brazil}

\date{\today}


\begin{abstract}
The combined effects of defect-defect interaction and of uniaxial or
biaxial strains of up to 10\% on the development of magnetic states on
the defect-core-localized quasi-one-dimensional electronic states
generated by the so-called 558 linear extended defect in graphene
monolayers are investigated by means of {\it ab initio}
calculations. Results are analyzed on the basis of the heuristics of
the Stoner criterion. We find that conditions for the emergence of
magnetic states on the 558 defect can be tuned by uniaxial tensile
parallel strains (along the defect direction) at both limits of
isolated and interacting 558 defects. Parallel strains are shown to
lead to two cooperative effects that favor the emergence of itinerant
magnetism: enhancement of the DOS of the resonant defect states
in the region of the Fermi level and tuning of the Fermi level to the
maximum of the related DOS peak. A perpendicular strain is likewise
shown to enhance the DOS of the defect states, but it also effects a
detunig of the Fermi level that shifts away from the maximum of the
DOS of the defect states, which inhibts the emergence of magnetic
states. As a result, under biaxial strains the stabilization of a
magnetic state depends on the relative magnitudes of the two
components of strain.
\end{abstract}

\maketitle

\section{Introduction}
\begin{figure}[b]
\includegraphics[width=8cm]{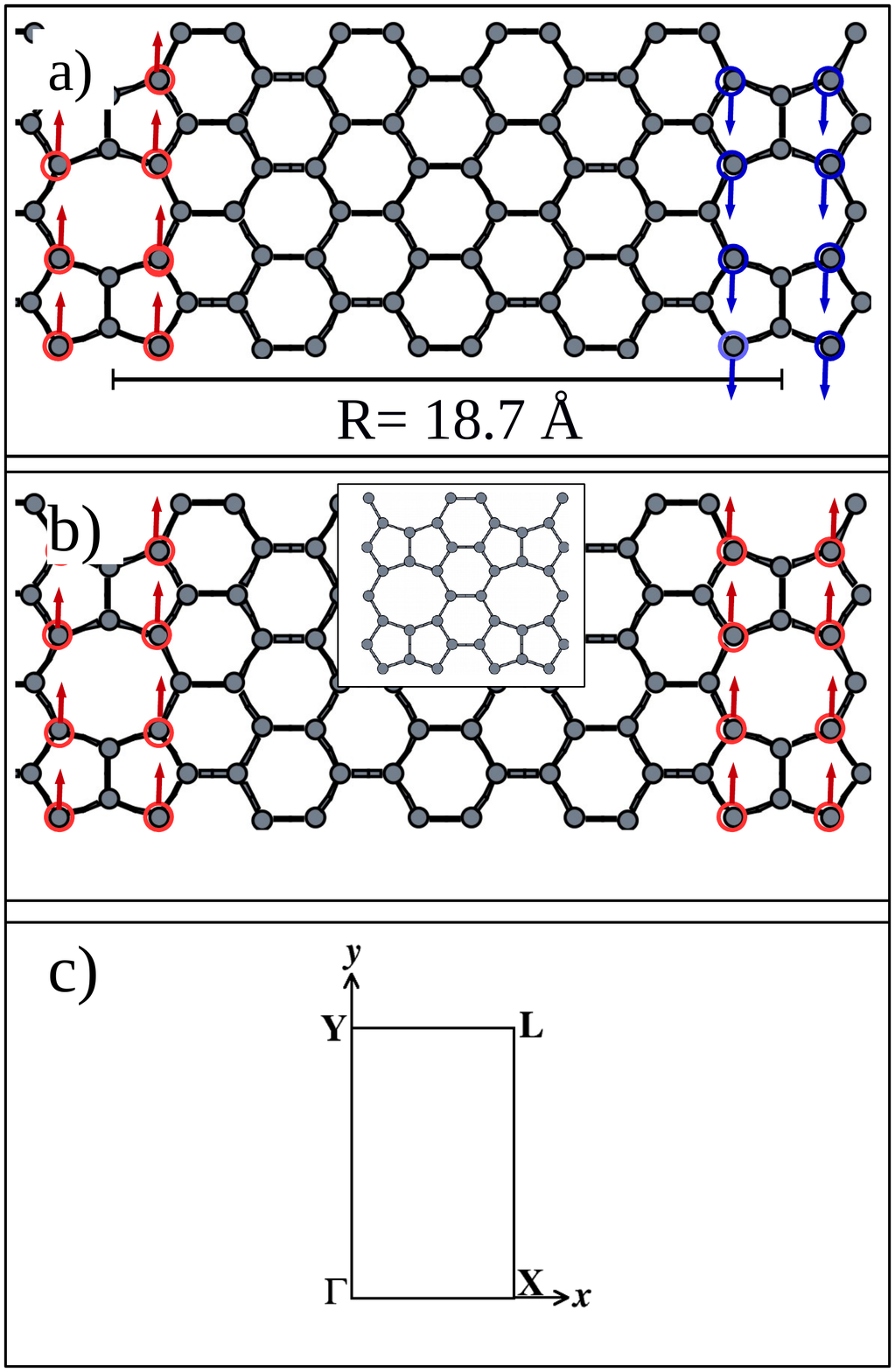}
\caption{(Color online) Geometry of the 558 extended defect supercells
  in graphene. (a) Shows two periods of the 557 defect and its nearest
  periodic image, for the case of a defect-defect distance of $R =
  18.7~\!$\AA. Initial arrangement of atomic spins for the
  antiferromagentic state is indicated. (b) Initial distribution of
  atomic spins for the ferromagnetic state. The inset shows a 2x2 cell
  of the case with $R = 5.7~\!$\AA. (c) Brillouin zone corresponding
  to the supercells in our study, showing special k-points along
  symmetry lines.}
\label{fig1}
\end{figure}

Technological applications of two-dimensional (2D) nanomaterials
require the ability to control their mechanical, electronic, and
magnetic properties. In the last decade, graphene in the 2D monolayer
form has become an important subject of research, motivated by its
mechanical strength and the rich electronic phenomenology connected
with the Dirac-fermion nature of its electronic structure near -
within a scale of $\sim$1~eV - the Fermi level.~\cite{rmp,geim,kats} The
origin of magnetism in graphene is not fully understood, being usually
associated with the presence of vacancies or adsorbates that tend to
bind to vacancies.~\cite{sepioni,wang,yazyev,elton,lehtinen} In bypartite
lattices, these defects lead to an imbalance in the electronic
occupation of the two sublattices, which leads to stabilization of
magnetic ground states, as predicted by the Lieb theorem~\cite{lieb}.
In a 2D material such as graphene, vacancies and topological point
defects can be created in non-equillibrium densities by electron-beam
irradiation~\cite{kotakoski}. However, full control over such magnetic
states is hampered by the random placement of vacancies.

Judicious introduction of structural defects presents an alternative
for manipulating the electronic and magnetic properties in 2D
materials~\cite{rmp,alexandre,ren,lahiri,td-ntwrk,vozmediano,white,joice,louiegb1,louiegb2,tsen}
Besides the tilt GBs that inevitably occur in polycrystalline
graphene~\cite{yu,simonis,cervenka,kim,huang}, a so-called 558
extended line defect was shown to occur in graphene layers grown on Ni
substrates~\cite{lahiri} as the interface across which the stacking of
the graphene layer with respect to the Ni substrate shifts from AB to
AC (in the usual convention for layer stacking in close-packed
lattices). Furthermore, recent experimental work has introduced a
protocol for the synthesis of this 558 extended defect in a
controllable fashion in a graphene monolayer~\cite{louie558}, which
shows that the possibility of manipulating the electronic and magnetic
properties of graphene and other two-dimensional materials, by
controllable introduction of defects, is a realistic prospect for the
near future.

The morphology of the 558 extended defect, shown in Fig.~\ref{fig1},
consists of a periodic unit composed of two side-sharing pentagonal
rings connected to an octagonal ring.  Alexandre {\it et
  al.}~\cite{alexandre} employed {\it ab initio} calculations to show
the development of itinerant ferromagnetism in the
quasi-one-dimensional (q1D) electronic states that are strongly
localized on the core of the 558 defect.  The ferromagnetic state
obtained by Alexandre {\it et al.} requires $n$-type doping in order
for the Stoner criterion for a ferromagnetic instability to be
met~\cite{stoner1,stoner2}. Electron doping shifts the Fermi level to the
maximum of the peak in the electronic density of states (DOS)
associated with the extended van Hove singularities that appear at and
near the Fermi level, that are characteristic of the q1D electronic
states generated by the 558 extended defect~\cite{alexandre}.

The extended van Hove singularities, related to the large flat
portions of the defect electronic states crossing the Fermi level,
signal a strong localization of the q1D defect states that leads to an
enhancement of exchange and correlation effects. Tuning the Fermi
level to the region of the maximum of the related DOS peak leads to
the onset of the magnetic states. Large periodic supercells were
employed in Ref.~\cite{alexandre}, with negligible couplings between
the 558 defect in the home cell and its periodic images, meaning that
conditions for the emergence of the magnetic state apply to the case
of an isolated and unstrained 558 defect in that study. One is
naturally led to consider the formation of magnetic states in the 558
defect in graphene under less restrictive conditions.

Two-dimensional materials grown on mismatched substrates are
commonly subject to strain. Moreover, strain engineering opens up the
possibility of tailoring electronic and magnetic functionalities in 2D
materials by the intentional application of strain. Graphene is known
to withstand strains as high as 20-25\% without
failure~\cite{clee,colombo,td-ntwrk}, being one of the 2D systems of choice for
strain engineering~\cite{ahcn}. Generally, tensile strains lead to
reduced band widths and extended van Hove singularities, and thus to
enhanced exchange and correlation effects, as exemplified by the case
of palladium, which displays paramagnetic states that are very close
to magnetic instabilities that can be triggered by quantum confinement
and strain in low-dimensional structures~\cite{paladio1,paladio2,alexandre3,alexandre4}.

In the present study, we employ {\it ab initio} calculations to
address the combined effects of defect-defect interaction and of
uniaxial or biaxial strains of up to 10\% on the development of
magnetic instabilities on the q1D electronic states generated by the
558 extended defect on graphene monolayers.

Our calculations indicate that conditions for the development of
magnetic instabilities on the core-localized q1D defect states can be
tuned by tensile uniaxial strains, along the defect direction, at both
limits of isolated and interacting 558 defects. A tensile strain along
the defect axis, which we refer to as a parallel strain, leads to two
cooperative effects that favor the emergence of itinerant
ferromagnetic in the 558-defect states: (i) enhancement of the DOS of
the q1D states in the region of the Fermi level and (ii) tuning of the
Fermi level to the maximum of the related DOS peak.

On the other hand, an uniaxial tensile strain in the direction
perpendicular to the defect line (perpendicular strain) is shown to be
detrimental to the development of magnetic states on the 558 defect,
because in this case, while we still obtain an enhancement of the DOS
of the q1D as in the case of a parallel strain, the Fermi level is
found to shift away from the maximum of the defect DOS, i.e., a
perpendicular strain leads to a detuning of the Fermi level that
inhibts the emergence of the magnetic states. As a result, under
biaxial strains we find that the stabilization of a magnetic state
depends on the relative magnitudes of the two components of strain,
parallel and perpendicular.

Regarding the meaning of our DFT-theory mean-field results, it must be
stressed that, because of their 1D nature, these correlated magnetic
states do not show long range order~\cite{mermim-wagner}. Instead,
they present algebraic correlation functions, and the magnetic states
we find in our calculations should manifest themselves in experimental
samples as magnetic domains with a null average macroscopic
magnetization.

\section{Methodology}

In our calculations we employ the SIESTA code~\cite{siesta}
implementation of Kohn-Sham density functional theory (DFT), within
the generalized-gradient approximation (GGA)~\cite{ks,gga} for the
exchange and correlation functional. Interactions between valence
electrons and ionic cores are treated using norm-conserving
pseudopotentials in the Kleinman-Bylander factorized
form~\cite{tm-kb1,tm-kb2}. A double-zeta LCAO basis set, augumented with
polarization orbitals, is used to expand the electronic wave
functions.  In all calculations, an equivalent real-space mesh cutoff
of 250 Ry is used, and meshes of up to 64 k-points along the
extended-defect direction are used to converge the electronic density
and the density of states.

Full structural relaxation is performed, with forces on atoms reaching
values of 0.01~eV/\AA\ or lower in all cases. For the equillibrium
(unstrained) geometries, the residual pressure on the supercell is
lower than 1~kBar in all cases. In a few selected cases, convergence
of energies and magnetic moments is verified with calculations
employing larger k-point sets and a mesh cutoff of 300 Ry, to ensure
that our results are converged with respect to calculational
parameters.

The supercells we employ, as shown in Fig.~\ref{fig1}, contain a
single 558 extended defect, and the supercell vector in the direction
perpendicular to the defect (the $x$-axis of the cell) determines the
nearest defect-defect distance $R$ in the periodic array of defects
generated by our use of periodic boundary conditions, as shown in
Fig.~\ref{fig1}. Supercells containing 558 defects may be classified
by the number $N$ of ``buffer'' zigzag chains of carbon atoms in the
bulk part of the cell, as suggested in Ref.~\cite{ren}.

In our analysis, we find it more expedient to classify the supercells
by the distance $R$ between the 558 defect in the home cell and its
closest periodic images. We consider a total of six different
supercells: $R = 5.7~\!$\AA\ (N = 0), $R = 10.0~\!$\AA\ (N = 2), $R =
14.3~\!$\AA\ (N = 4), $R = 18.7~\!$\AA\ (N = 6), $R = 23.0~\!$\AA\ (N = 8), and
$R = 27.3~\!$\AA\ (N = 10). These supercells cover the range of
defect-defect distances between $R = 5.7~\!$\AA, the smallest possible
distance between adjacent 558 defects, and $R = 27.3~\!$\AA, a value at
which defect-defect interaction is neglible and the electronic
properties of the defect are characteristic of isolated defects.

The geometry of the supercell with $R = 18.7~\!$\AA, with six
buffer chains between defects, is shown in Fig.~\ref{fig1}. The inset
in Fig.~\ref{fig1}(b) shows a 2x2 frame of the supercell with $R =
5.7~\!$\AA, with no buffer chains between the 558 defect in the home
cell and its closest periodic images.

\section{Results and Discussion}

Our results are analyzed in terms of the Stoner criterion (SC) for
itinerant magnetic instabilities:
\begin{equation}
I{\cal N}\left(\epsilon_F\right) \ge 1\;\;;
\label{eq1}
\end{equation}
where $I$ is the exchange integral and ${\cal  N}\left(\epsilon_F\right)$
is the DOS at the Fermi level. [We denote the DOS at energy $\epsilon$ as
${\cal N}\left(\epsilon\right)$.]

Our focus is to address the effects of defect-defect interaction and
strain on the development of magnetic states on the q1D electronic
states of the 558 defect in graphene, based on the heuristics of the
SC. While the strong localization of the q1D defect states favors both
factors in the left-hand side of the SC inequality, in
Ref.~\onlinecite{alexandre} it was shown that, in the isolated-defect
limit, tuning the Fermi level with $n$-type doping is required for the
ferromagnetic instability to set in, which means that the SC is not
met for an isolated 558 defect in a neutral and unstrained graphene
layer.

We consider ferromagnetic (FM) and antiferromagnetic (AFM) couplings
between defects~\cite{ren}, as well as the spin-unpolarized
non-magnetic case (NM). Figure~\ref{fig1} shows schematically the
starting spin distribution for the initial states of the FM and AFM
states in our DFT calculations. After electronic self-consistency is
achieved, we obtain the corresponding FM and AFM states for the 558
defect.  We have also attempted several other initial spin
configurations, such as an antiferromagnetic coupling between the two
zigzag chains on the core of the 558 defect (as considered in
Ref.~\cite{ren}), as well as other initial antiferromagnetic
arrangements of initial spin states for the atoms along the core of
the 558 defect. At the GGA level, these converge either to the FM or
to the AFM states shown in Fig.~\ref{fig1}.

In our calculations, we have imposed isotropic and anisotropic biaxial
strains as well as parallel and perpendicular uniaxial strains.  In
the following discussion we concentrate our analysis first on the
effects of tensile parallel strains, followed by a discussion on the
effects of tensile perpendicular strains and biaxial strains.

\subsection{Energetics}
\begin{figure}[b]
\includegraphics[width=8cm]{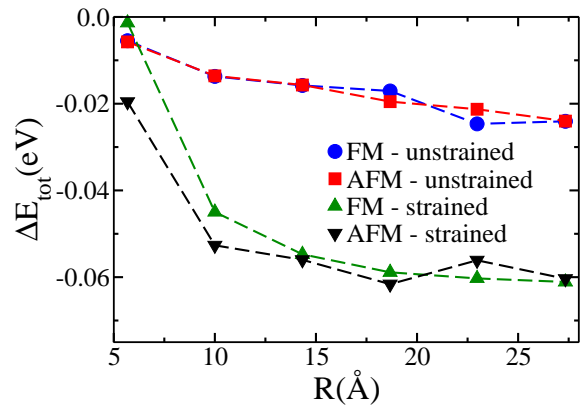}
\caption{(Color online) Relative energies of magnetic and non-magnetic
  states of the 558 extended defect (see text). Blue circles show
  energy of the ferromagnetic (FM) state of the unstrained defect,
  relative to the non-magnetic case, as a function of defect-defect
  distance. Red squares show the same for the antiferromagnetic (AFM)
  unstrained state. Green up triangles show energy of the FM state at
  a 10\% uniaxial tensile strain (along the defect direction). Black down
  triangles show the same for the AFM state.}
\label{fig2}
\end{figure}

We start by addressing the combined effects of defect-defect
interaction and a homogeneous parallel strain on the energetics and
magnetic states of the defect in the neutral (undoped)
case. Figure~\ref{fig2} shows the difference in total energy per
defect periodic unit, with respect to the energy of the NM state, for
the FM ($\Delta E_{tot}^{FM} = E_{tot}^{FM} - E_{tot}^{NM}$) and AFM
states ($\Delta E_{tot}^{AFM} = E_{tot}^{AFM} - E_{tot}^{NM}$) of the
558 defect as functions of the defect-defect distance $R$. The figure
shows $\Delta E_{tot}$ for the equillibrium (non-strained) as well as
for the uniaxially-strained cases, at a parallel strain of 10\%.

Figure~\ref{fig2} shows that the equillibrium FM and AFM states are
nearly energy-degenerate for all values of $R$. A discernable trend is
that, for both the FM and AFM states, $\Delta E_{tot}$ increases in
magnitude as the defect-defect interaction is reduced with increasing
$R$. A small energy difference of $\sim$2.5~meV (per defect periodic
unit), favoring the AFM state is observed for $R = 18.7~\!$\AA, while
for $R = 23.0~\!$\AA\ the FM state is favored by $\sim$3.4~meV. At the
largest distance of $R = 27.3~\!$\AA, defect-defect interaction effects
are negligible and the two phases are degenerate, with $E_{tot}$
values that are smaller than the NM case by 24 meV per defect periodic
unit.

The energetics of the magnetic states of the strained 558 defect shows
a richer structure. The AFM state is favored for all values of $R$,
except for the case of $R = 23.0~\!$\AA, and the FM-AFM split in energy
is much larger at small defect-defect separations than in the
unstrained case, with the AFM state being favored by 20~meV at the
smaller distance of $R = 5.7~\!$\AA. At larger defect-defect distances,
the FM and AFM states become nearly degenerated, with energies that
are lower than the NM case by $\sim$55-62~meV, compared with the
unstrained results of 24~meV. Generally, a parallel uniaxial strain
enhances the stability of the magnetic states with respect to the NM
state.

\subsection{Magnetic States: Effects of Tensile Parallel Strain}

\begin{figure}[b]
\includegraphics[width=8cm]{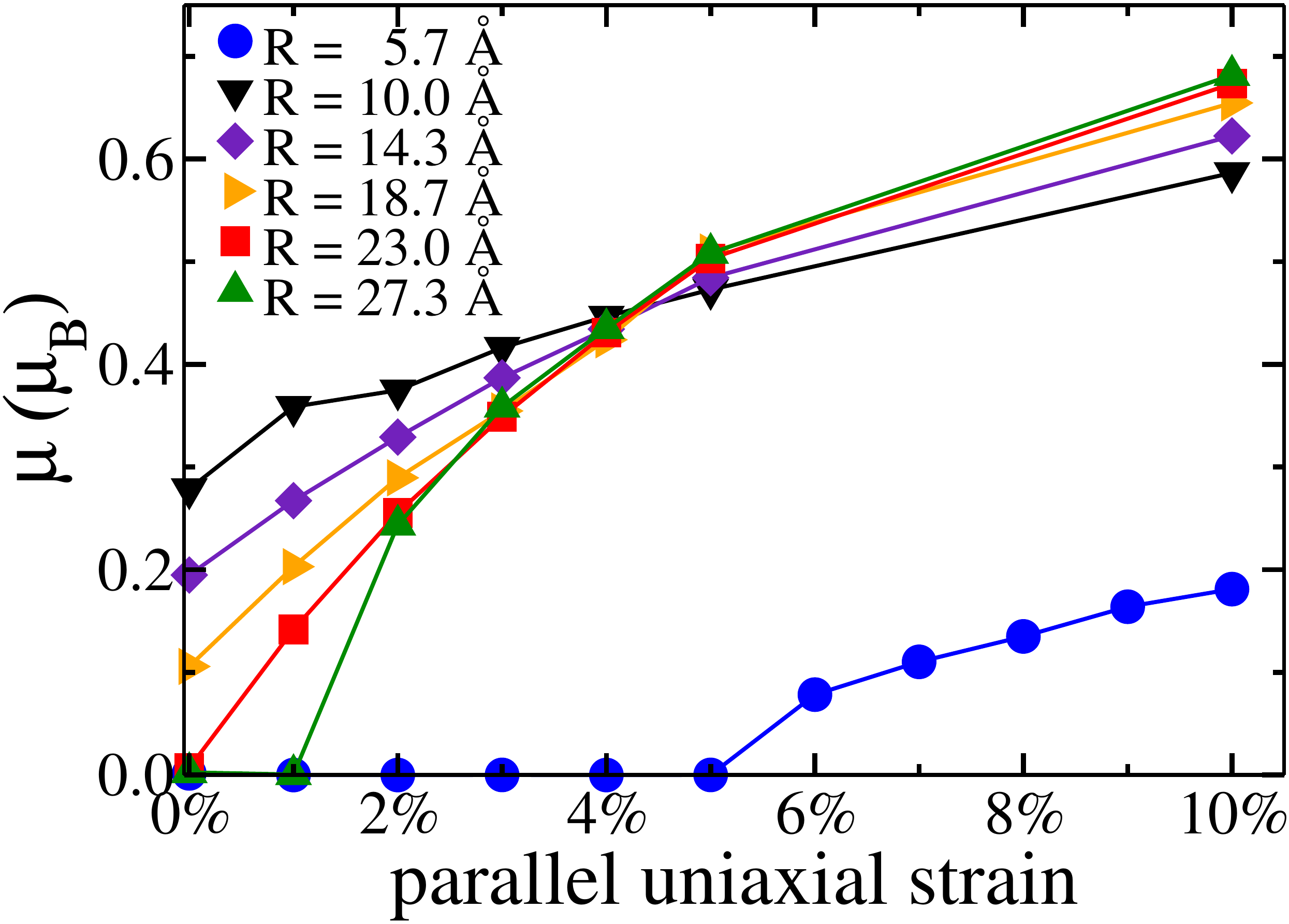}
\caption{(Color online) Values of magnetic moment per defect periodic
  unit, $\mu$ in units of the Bohr magneton, as a function of a
  parallel uniaxial tensile strain (see text), for the six values of
  defect-defect distances.}
\label{fig3}
\end{figure}

Shifting gears now to the onset of magnetic states, we show in
Fig.~\ref{fig3} the magnetic moment per defect unit $\mu$ (in units of
the Bohr magneton, $\mu_B$) as a function of uniaxial tensile parallel
strain for different values of $R$. For the unstrained cases (0\%
strain), we observe that $\mu$ decreases with increasing $R$, with the
exception of the anomalous case of $R = 5.7~\!$\AA, that shows a very
small value of $\mu$.  Magnetic moment values at the largest
defect-defect separations in our study, $\mu = 0.007~\mu_B$ for
$R = 23.0~\!$\AA\ and $\mu = 0.003~\mu_B$ for $R = 27.3~\!$\AA, are very
small for the unstrained defects. Figure~\ref{fig3} also shows that
the rate of increase of $\mu$ with strain increases with $R$.

Indeed, at a 4\% parallel strain the values of $\mu$ are nearly the
same for all values of $R$ (with the exception of the anomalous case
of $R = 5.7~\!$\AA\ that we discuss in more detail below), and for a
strain of 10\% the behavior of $\mu$ as a function of $R$ is reversed,
and the magnetic moment becomes an increasing function of the
defect-defect separation, for the range of $R$ values we
consider. Note that for the larger defect-defect distances ($R =
23.0~\!$\AA\ and $R = 27.3~\!$\AA ), for strains between 1\% and 2\% the
$\mu$ values increase by two orders of magnitude and become comparable
to those for smaller values of $R$.  The case of $R = 5.7~\!$\AA\ is
anomalous, with very small values of $\mu$ for strains up to 5\%. In
this case, strains larger than a critical value between 5\% and 6\%
are needed for $\mu$ to reach values of 0.1-0.2~$\mu_B$.
\begin{figure}[b]
\includegraphics[width=8cm]{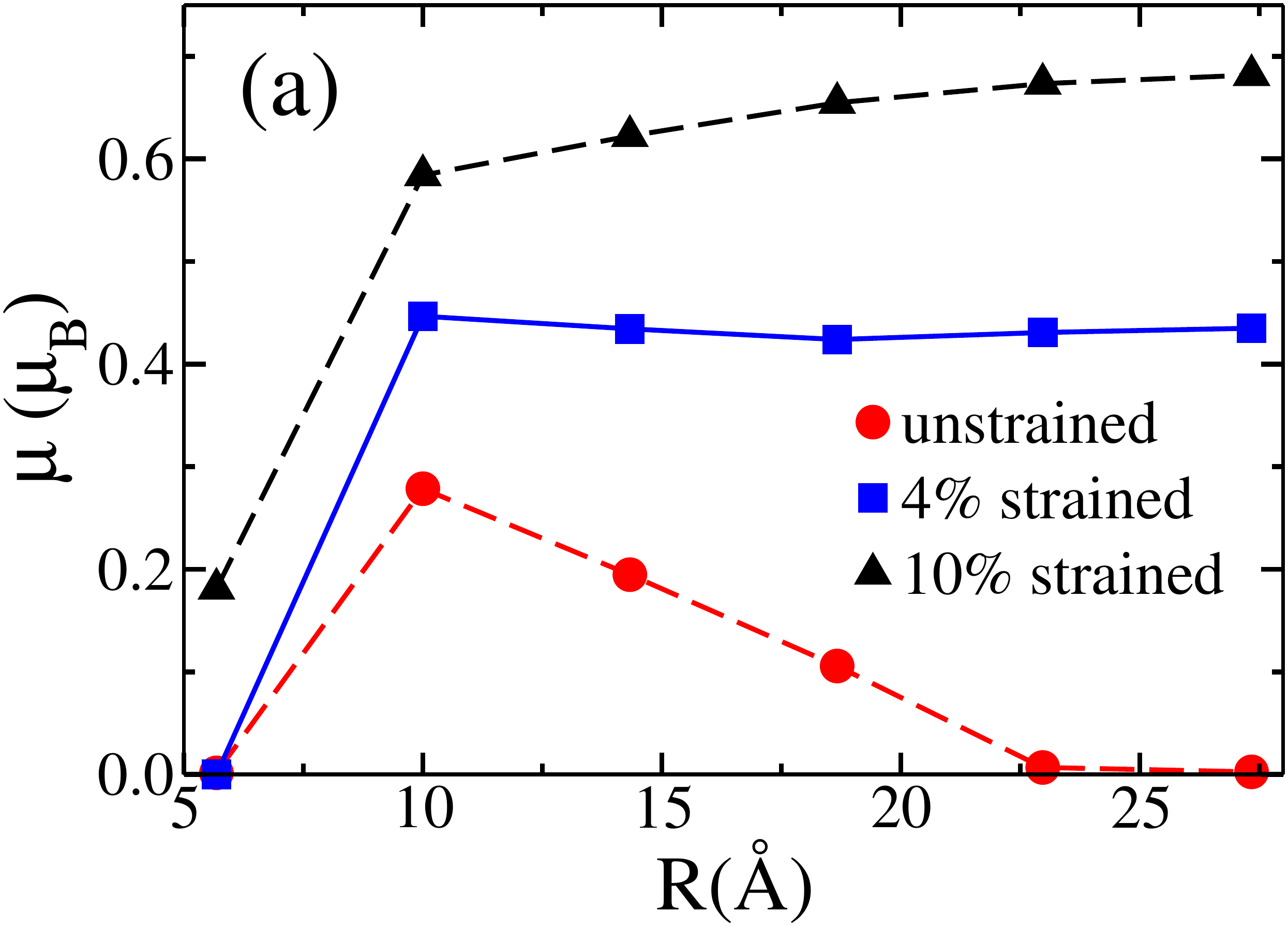}
\caption{(Color online) Magnetic moments as a function of
  defect-defect distance, for the unstrained, and for parallel tensile strain
  values of 4\% and 10\%.}
\label{fig4}
\end{figure}

In order to facilitate the visualization of these trends, in
Fig.~\ref{fig4} we plot the data from Fig.~\ref{fig3} as a function of
$R$ for three different values of tensile parallel strain: unstrained,
4\%, and 10\%. Generally, from Figs.~\ref{fig3} and \ref{fig4} we
conclude that defect-defect interaction favors the emergence of
itinerant magnetism in unstrained 558 defects, with the exception
of the case of defects at their closest possible separation ($R = 5.7~\!$\AA).
In its turn, a tensile parallel strain also leads to the onset of
magnetism, and supersedes the effect of interaction, starting at about
a 4\% parallel strain, as shown in Fig.~\ref{fig4}. At 4\% parallel
strain, the magnetic moment per defect unit is nearly independent of
$R$ (for $R\ge 10.0~\!$\AA), and for larger strains $\mu$ increases with
distance, in stark contrast with the behavior of the unstrained
defects.
\begin{figure}[b]
\includegraphics[width=8cm]{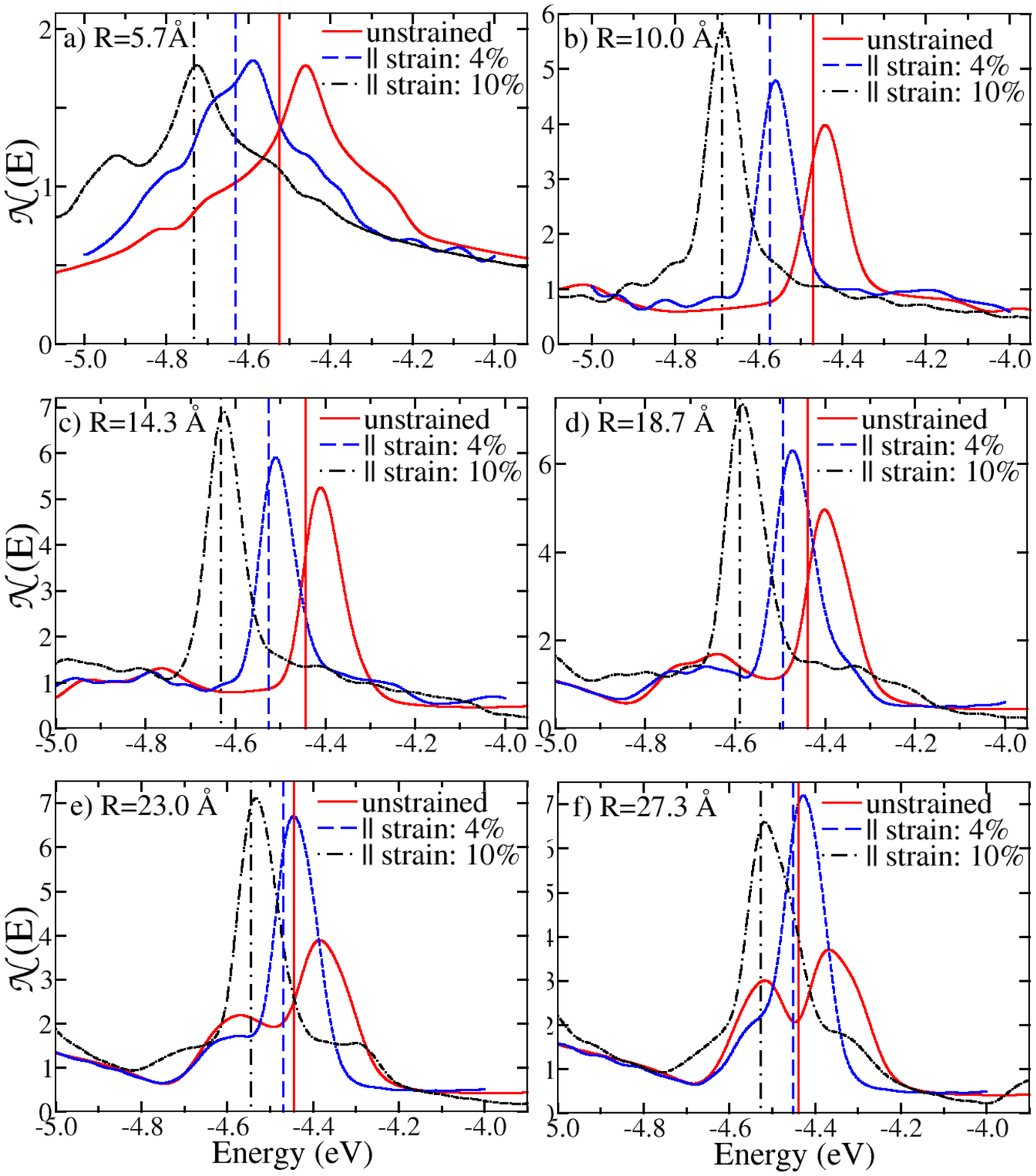}
\caption{(Color online) Evolution of the density of states ${\cal
    N}(\epsilon)$ with strain for the six values of $R$, showing the
  curves for three different values (unstrained, 4\%, and 10\%) of
  parallel uniaxial tensile strain.}
\label{fig5}
\end{figure}

The mechanisms behind these trends, and also behind the behavior of
the anomalous case of defects at a distance of $R = 5.7~\!$\AA, can be
understood from the perspective of the Stoner
criterion. Figure~\ref{fig5} shows the DOS, ${\cal
  N}\left(\epsilon\right)$, in the Fermi level region, for the six
defect-defect distances we consider. For each case, we show ${\cal
  N}\left(\epsilon\right)$ for the unstrained state, and for the cases
of 4\% and 10\% parallel strain.

Starting from the anomalous $R = 5.7~\!$\AA\ case in
Fig.~\ref{fig5}(a), we observe that application of a tensile strain
does not enhance the defect-related peak in the DOS near the Fermi
level. This indicates that strain does not enhance the localization of
the defect electronic states at the Fermi level, and its only effect
in this case is a better tuning of the Fermi level, that shifts closer
to the maximum of the DOS peak in Fig.~\ref{fig5}(a). This explains
why, at this defect-defect distance, the 558 defect only develops a
magnetic moment for strains above 5\%. Thus, the only factor in
Eq.~\ref{eq1} that is affected by application of a tensile parallel
strain in this case is the value of the DOS at the Fermi level, ${\cal
  N}\left(\epsilon_F\right)$, and the exchange integral remains
essentially unchanged.
\begin{figure}[b]
\includegraphics[width=8cm]{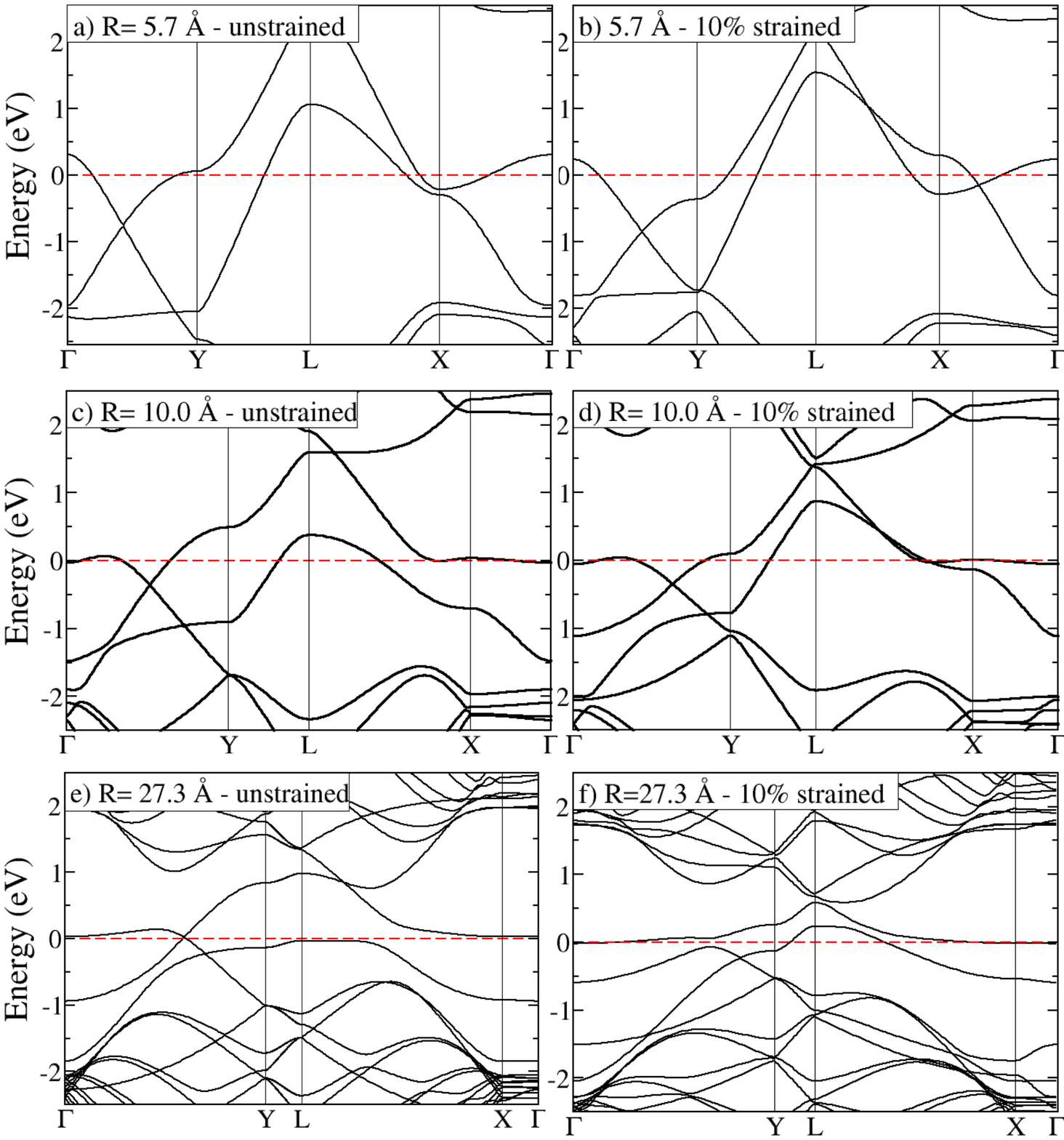}
\caption{(Color online) Evolution of the band structure of defect
  states with respect to parallel tensile strain. (a) and (b) show the
  band structures of the strained and 10\%-strained cases,
  respectively, for a defect-defect distance of $R = 5.7~\!$\AA. (c)
  and (d) show the same for $R = 10.0~\!$\AA. (e) and (f) show the
  same for $R = 27.3~\!$\AA.}
\label{fig6}
\end{figure}

At larger distances this picture changes, as shown in
Figs.~\ref{fig5}(b)-(f) where we observe that a parallel tensile
strain induces two effects in the electronic structure of the 558
defect. The first one is a better tuning of the Fermi level, that
shifts closer to the peak of defect states in the DOS, as in the $R =
5.7~\!$\AA\ case, but we also observe an enhancement of the height of
the DOS peak, connected with a stronger localization of the electronic
states of the 558 defect in the Fermi level region. Note that at the
two largest distances in our study ($R = 23.0~\!$\AA\ and $R =
27.3~\!$\AA), the evolution of the DOS and also of the Fermi level
position with strain is more complex due to the presence of two DOS
peaks of defect states near the Fermi level in the unstrained state,
that merge into a single peak at larger strains.

Regarding the anomalous $R = 5.7~\!$\AA\ case, we speculate that when
558 defects are separated by a distance $R = 5.7~\!$\AA, the lack of a
bulk region (as shown in the inset in Fig.~\ref{fig1}) onto which the
defect-related electronic states can relax, inhibits the enhancement
of the localization of the defect states induced by the parallel
strain, that we observe at larger distances where the defects are
surrounded by bulk material.

The parallel-strain induced enhancement of localization, hence of the
exchange integrals in Eq.~\ref{eq1}, is connected with the changes in
the 558-defect electronic states. In Fig.~\ref{fig6}, we show the band
structures for the unstrained and 10\% parallel-strained cases, for
three different values of $R$. Figure~\ref{fig6}(a) shows the
$R=5.7~\!$\AA\ case, where extended van Hove singularities do not appear
in the band structure of the unstrained defect. At this defect-defect
distance, even for the 10\%-strained case in Fig.~\ref{fig6}(b) we
observe no extended van Hove singularities in the band strucuture, and
the only effect of strain is the tuning of the Fermi level, as
discussed in the previous paragraph. This explains why for
$R=5.7~\!$\AA\ the system develops only a moderate value of $\mu$, even
for such large value of strain, and generally the behavior of $\mu$
with strain does not follows the trends we observe at larger
defect-defect distances.

When 558 defects are separated by $R = 10.0~\!$\AA, extended van Hove
singularities, connected to the large flat portions of the defects
bands at the Fermi level, appear in the band structure for the
unstrained defect, as shown in Fig.~\ref{fig6}(c).  This observation,
along with the fact that the Fermi level in this case is very near the
maximum of the DOS peak, are the reasons behind the development of a
large value of $\mu$ already for the unstrained defects in this
case. For the strained states, a better tuning of the Fermi level to
the maximum of the resonant defect peak in the DOS [shown in
  Fig.~\ref{fig5}(b)], coupled with an enhancement of the extended van
Hove singularities, due to an elargement of the flat portions of the
bands at the Fermi Level, and a reduction of the band widths of the
defect bands, shown in Figs.~\ref{fig6}(d), explain the increase of
$\mu$ with increasing strain displayed in Fig.~\ref{fig3}.

Strain plays an even more decisive role in the limit of non-interacting
defects ($R = 27.3~\!$\AA). As indicated above, the unstrained defect
shows a borderline behavior, with a very small value of $\mu$. Given
the presence of quite wide van Hove singularities in the band
structure of the unstrained defect, as displayed in
Fig.~\ref{fig6}(c), it is to be expected that tensile strains in this
case should drive the system towards a more robust magnetic
state. Indeed, for a critical parallel tensile strain between 1\% and
2\% the system develops a sizeable value of $\mu = 0.24~\mu_B$.
Further increase in the value of $\mu$ for larger strains is explained
along the same reasoning as the $R = 10.0~\!$\AA\ case, i.e., better
tuning of the Fermi level and an enhancement of the flat portions of
the defect bands that leads to more localized states and enhanced
exchange effects. For this non-interacting case, Fig.~\ref{fig6}(f)
shows a marked increase in the extent of the extended van Hove
singularities lying at the Fermi level, at 10\% parallel strain, along
the ${\bf \Gamma-}${\bf Y} and {\bf L-X} lines in the Brillouin zone
[both are parallel to the defect direction, as shown in
  Fig.~\ref{fig1}(c)].

\subsection{Magnetic States: Effects of Perpendicular and Biaxial Tensile Strains}

The above discussion shows that a parallel tensile strain favors the
emergence of magnetic states in the 558 defect in graphene, by
enhancing both factors encoded in the Stoner criterion in
Eq.~\ref{eq1}. In the present section, we consider the effects of
biaxial and perpendicular tensile strains, analyzing in detail the
strongly interacting $R = 10.0~\!$\AA\ case, that displays the largest
value of $\mu$ for the unstrained defects, and the non-interacting $R
= 27.3~\!$\AA\ case. These two examples suffice to highlight the general
trends and the generality of the results will be pointed out as we
proceed with the discussion.

\subsubsection{Perpendicular Tensile Strain}

\begin{figure}[b]
\includegraphics[width=8cm]{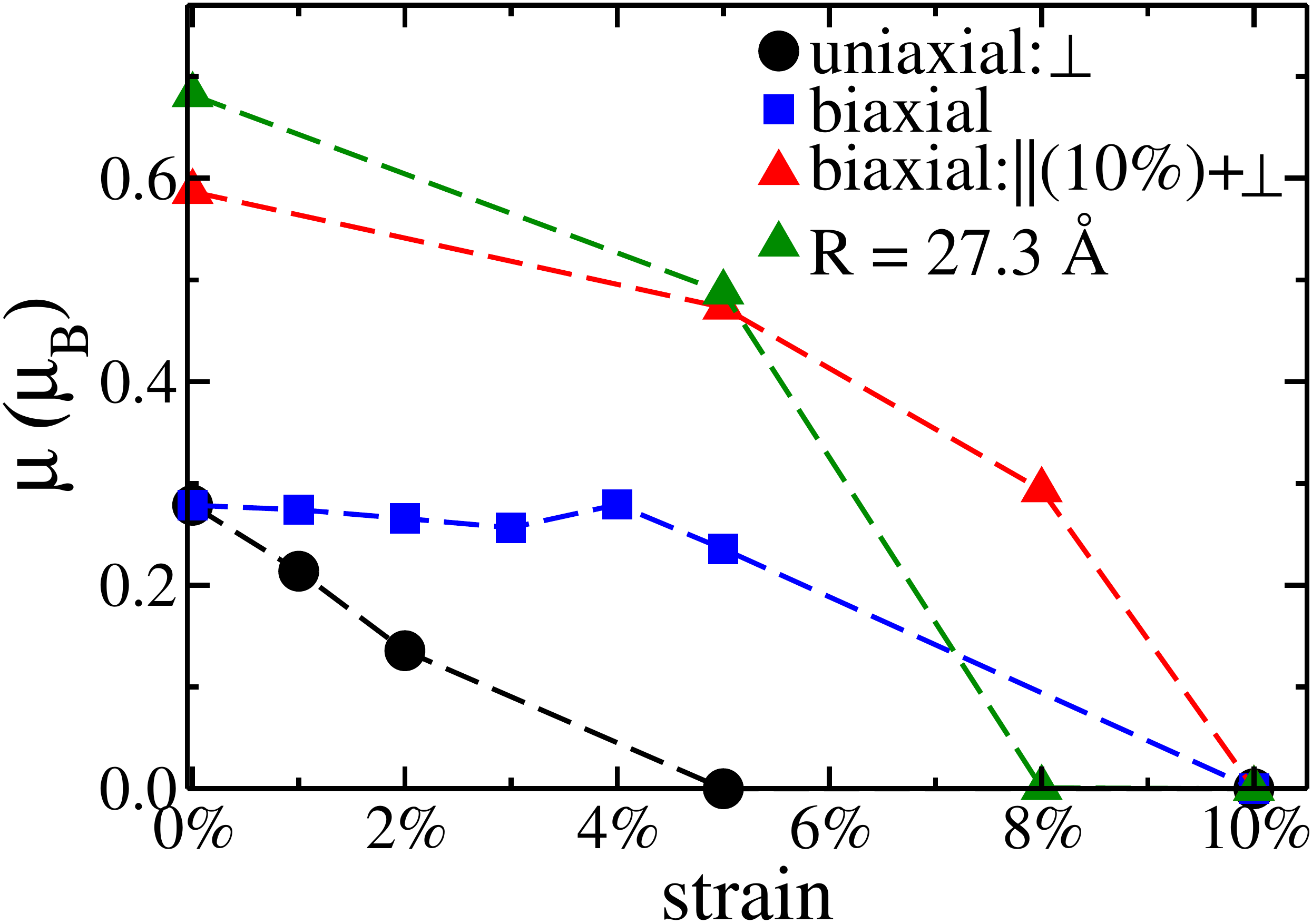}
\caption{(Color online) Magnetic moment as a function of perpendicular
  uniaxial strain and of isotropic and anisotropic biaxial
  strains. Black circles show $\mu$ as a function of perpendicular
  tensile strain for the $R = 10.0~\!$\AA\ case. Blue squares show
  $\mu$ as a function of isotropic biaxial tensile strain for the $R =
  10.0~\!$\AA\ case. Red triangles show $\mu$ as a function of the
  perpendicular component of an anisotropic biaxial tensile strain for
  the $R = 10.0~\!$\AA\ case, with a fixed value of 10\% for the
  parallel component of strain. Green triangles show $\mu$ as a
  function of the perpendicular component of an anisotropic biaxial
  tensile strain for the $R = 27.3~\!$\AA\ case, with a fixed value of
  10\% for the parallel component of strain. }
\label{fig7}
\end{figure}

Filled black circles in Fig.~\ref{fig7} show the effect of
perpendicular uniaxial tensile strains from 1\% to 5\% on the magnetic
states of the $R = 10.0~\!$\AA\ case.  At a 5\% strain the magnetic
moment of the defect states has dropped to zero, indicating at a first
sight that a perpendicular tensile train does not favor the emergence
of magnetic states in the 558 defect. The issue, however, calls for a
more detailed analysis.

\begin{figure}[b]
\includegraphics[width=8cm]{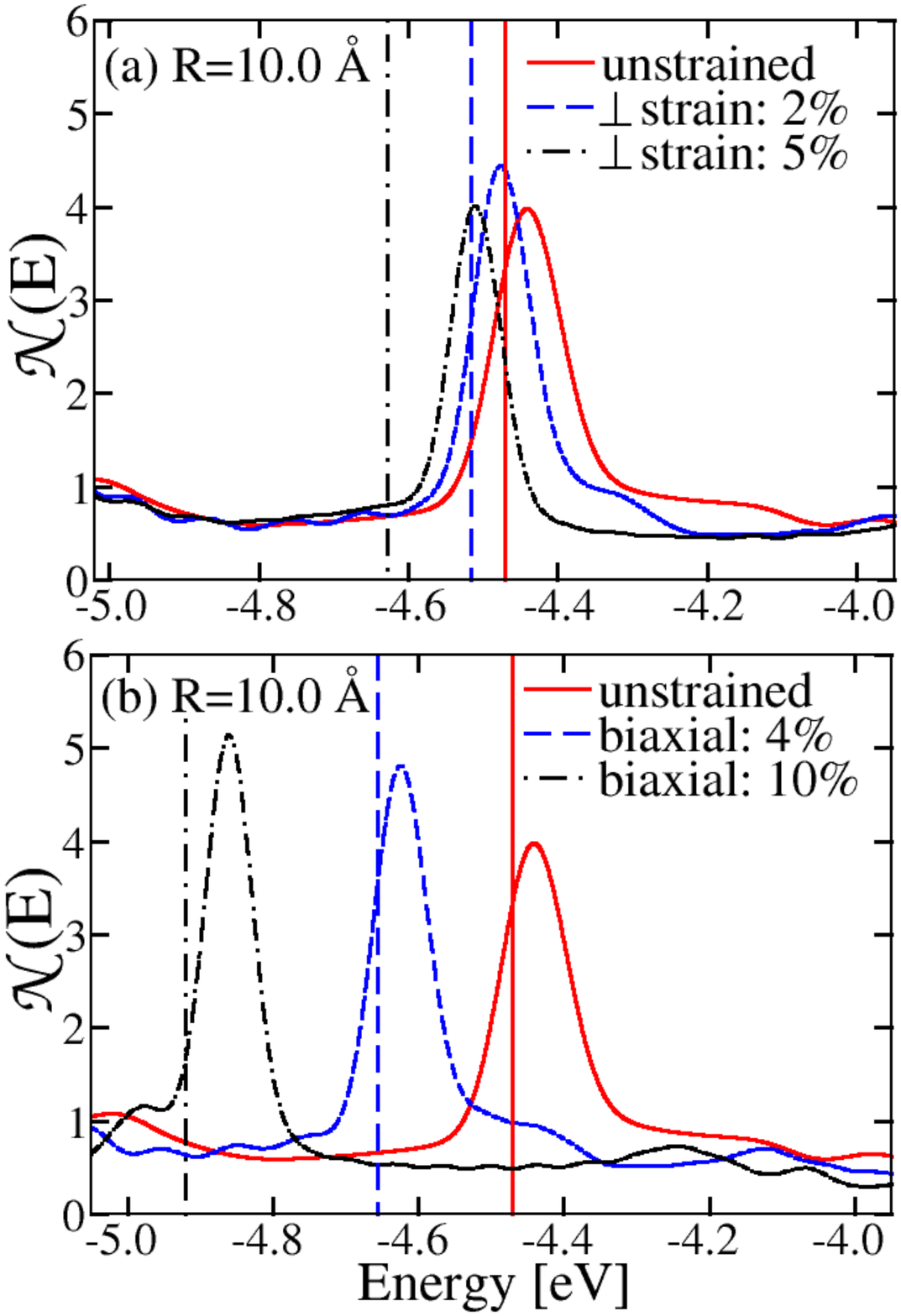}
\caption{(Color online) (a) Evolution of the density of states with
  respect to a perpendicular tensile strain for $R = 10.0~\!$\AA.  (b)
  Evolution of the density of states with respect to isotropic tensile
  biaxial strain for $R = 10.0~\!$\AA.}
\label{fig8}
\end{figure}

Figure~\ref{fig8}(a) shows the evolution of ${\cal N}(\epsilon)$ with
strain for this case. The first observation to be drawn from
Fig.~\ref{fig8}(a) is that a tensile perpendicular of up to 5\%
enhances the peak related to the defect states in the DOS, i.e., a
perpendicular strain may actually enhance the localization and thus the
exchange integral of the defect states. However, the figure also shows
a fast detuning of the Fermi level, connected with the overall changes
in the band structure induced by the application of the perpendicular
strain. Note that the height of the defect peak in ${\cal N}(\epsilon)$,
at 2\% perpendicular strain, is greater than that of the unstrained
defect, while the value of $\mu$ decreases in Fig.~\ref{fig7}, the
reason being the smaller value of ${\cal N}(\epsilon_F)$ that results
from the shift in the position of the Fermi level.

Therefore, quenching of the magnetic moment in this case is connected
to a low value of ${\cal N}(\epsilon_F)$, and not to a delocalization
effect that would impact the exchange integral. This is an important
observation, because the Fermi level can be tuned by doping, and we
can anticipate that magnetic states in graphene samples under
perpendicular tensile strains, containing 558 defects, may be induced
by proper Fermi-level tuning.

\subsubsection{Biaxial Strain}

Now that we have analyzed the effects of parallel and perpendicular
uniaxial tensile strains, we conclude by addressing the effects of
biaxial tensile strains. From the foregoing discussion, we know that
both parallel and perpendicular strains tend to enhance the extended
van Hove singularities of the defect states, hence both lead to an
enhancement of the associated exchange integrals. On the other hand,
they produce opposite effects in ${\cal N}(\epsilon_F)$, the value of
the DOS at the Fermi level, with parallel strains of up to 10\%
leading to larger values of ${\cal N}(\epsilon_F)$, while
perpendicular strains lead to a fast detuning of the Fermi level and
hence to rather low values of ${\cal N}(\epsilon_F)$.

We have performed calculations of isotropic biaxial strains of up to
10\% for all defect-defect distances in our study. We obtain a
ferromagnetic state under isotropic biaxial strain only for the $R =
10.0~\!$\AA\ case. For the other five values of $R$, magnetic moments
were either null or negligible for all values of biaxial strain from
1\% to 10\%.

The behavior of $\mu$ with strain for the $R = 10.0~\!$\AA\ case is
shown by the filled squares in Fig.~\ref{fig7}. We obtain that for
biaxial strains of up to 5\% the tuning and detuning effects of the
two components of strain nearly cancel each other, and $\mu$ remains
nearly constant, but a downwards trend can be observed already for
strains between 4\% and 5\%. For a 10\% tensile biaxial strain, $\mu$
vanishes, due to the Fermi-level detuning associated with the
perpendicular component of strain. In Fig.~\ref{fig8}(b) we show the
evolution of the DOS with biaxial strain for this case.

We have also addressed the case of nonisotropic biaxial strains.
Filled red triangles in Fig.~\ref{fig7} show the combined effects of a
10\% parallel strain coupled with perpendicular tensile strains
of 5\%, 8\%, and 10\%, for the $R = 10.0~\!$\AA\ and $R = 27.3~\!$\AA\
cases. For both values of $R$, a homogeneous biaxial strain of 10\%
produces an unpolarized spin state, due to a rather low value of
${\cal N}(\epsilon_F)$, as shown for $R = 10.0~\!$\AA\ in
Fig.~\ref{fig8}(b). Note that a substantial enhancement of the
defect-related peak in ${\cal N}(\epsilon)$, with respect to the
unstrained state, is observed, but the value of ${\cal N}(\epsilon_F)$
is rather small due to the Fermi-level detuning effect prompted by the
perpendicular component of strain. These observations also apply to the
$R = 27.3~\!$\AA\ case.

For smaller values of perpendicular strains, we obtain that: (i) for
the more robust magnetic state of the $R = 10.0~\!$\AA\ case, $\mu
=$~0.47~$\mu_B$ for a 5\% perpendicular strain and $\mu
=$~0.29~$\mu_B$ for an 8\% perpendicular strain, both combined with a
10\% parallel strain. For the non-interacting defects at $R =
27.3~\!$\AA, full quenching of the magnetic moment is observed when the
parallel strain is combined with an 8\% perpendicular strain, and $\mu
=$~0.49~$\mu_B$ for a 5\% perpendicular strain. As with the case of
isotropic biaxial strains, competition between the Fermi-level tuning
and detunig effects, due respectively to the parallel and
perpendicular components of strain determine the fate of the magnetic
states of the 558 defect under anisotropic biaxial strains.

\subsection{Conclusions}

In this work, we have addressed the combined effects of defect-defect
interaction and of uniaxial or biaxial strains of up to 10\% on the
development of magnetic instabilities on the quasi-one-dimensional
(q1D) electronic states generated by the 558 extended defect on
graphene monolayers, by means of {\it ab initio} calculations. We have
considered uniaxial strains along the defect direction (parallel
strain) and along the perpendicular direction (perpendicular strain),
and isotropic and anisotropic biaxial strains. We frame our results on
the basis of the Stoner criterion for itinerant magnetism, and analyze
the effects of the various strain states on the basis of their impact
on the two ingredients encoded in the Stoner criterion: localization
of the defect-generated electronic states in the region of the Fermi
level and the magnitude of the DOS of the defect states at the Fermi
level.

We obtain that conditions for the development of magnetic
instabilities in the defect states can be tuned by tensile
uniaxial parallel strains at both limits of isolated and interacting
558 defects. Parallel strains are shown to lead to two cooperative
effects that favor the emergence of itinerant magnetism in the
558-defect states: enhancement of the DOS of the defect states in the
region of the Fermi level and tuning of the Fermi level to the maximum
of the related DOS peak.

On the other hand, a perpendicular strain is shown to effect an
enhancement of the DOS of the defect states, as in the case of a
parallel strain, but the Fermi level is found to shift away from the
maximum of the DOS of the resonant defect states, i.e., a
perpendicular strain leads to a detuning of the Fermi level that
inhibts the emergence of the magnetic states.

Given the tuning and detunig of the Fermi level promoted,
respectively, by parallel and perpendicular uniaxial strains, under
biaxial strains we find that the stabilization of a magnetic state
depends on the relative magnitudes of the two components of strain.

\textbf{ACKNOWLEDGMENTS}

We acknowledge support from Brazilian agencies CNPq, FAPEMIG, Rede de
Pesquisa em Nanotubos de Carbono, INCT de Nanomateriais de Carbono,
and Instituto do Mil\^enio em Nanotecnologia-MCT.


\begin{thebibliography}{99}

\bibitem{rmp} A.~H.~Castro Neto, F.~Guinea, N.~M.~R.~Peres, K.~S.~Novoselov,
and A.~K.~Geim, \newblock{\em Rev.~Mod.~Phys.} {\bf 81}, 109 (2009) and references therein.

\bibitem{geim} A. K. Geim and K. S. Novoselov, \newblock{\em Nat. Mater.} {\bf 6}, 183 (2007).

\bibitem{kats} M.~I.~Katsnelson, K.~S.~Novoselov, and A.~K.~Geim,
\newblock {\em Nat. Phys.} {\bf 2}, 620 (2006).

\bibitem{sepioni} M. Sepioni, R. R. Nair, S. Rablen, J. Narayanan, 
F. Tuna, R. Winpenny, A. K. Geim, and I. V. Grigorieva, \newblock{\em Phys. Rev. Lett.}
{\bf 105}, 207205 (2010).

\bibitem{wang} Y.~Wang, Y.~Huang, Y.~Song, X.~Zhang, Y.~Ma, J.~Liang, and Y.~Chen,
\newblock{\em Nano Lett.} {\bf 9}, 220 (2009).

\bibitem{yazyev} O.~V.~Yazyev and L.~Helm, \newblock{\em Phys. Rev. B}
{\bf 75}, 125408 (2007).

\bibitem{elton} E.~J.~G.~Santos, A.~Ayuela, and D.~S\'anchez-Portal,
\newblock{\em New Journal of Physics} {\bf 14}, 043022 (2012).

\bibitem{lehtinen} P.~O.~Lehtinen, A.~S.~Foster, Y.~Ma, A.~V.~Krasheninnikov,
and R.~M.~Nieminen,
\newblock {\em Phys.~Rev.~Lett.} {\bf 93}, 187202, (2004).

\bibitem{lieb} E.~H.~Lieb \newblock{\em Phys. Rev. Lett.} {\bf 62}, 1201 (1989).

\bibitem{kotakoski} J.~Kotakoski, A.~V.~Krasheninnikov, U.~Kaiser, and J.~C.~Meyer,
\newblock {\em Phys. Rev. Lett.} {\bf 106}, 105505 (2011).

\bibitem{alexandre} S.~S.~Alexandre, A.~D. L\'ucio and A.~H.~ Castro Neto and R.~W.~Nunes,
\newblock {\em Nano Lett.} {\bf 12}, 5097 (2012).

\bibitem{ren} [46] J.-C.~Ren, Z.~Ding, R.-Q.~Zhang, and M.~A.~Van Hove,
\newblock{\em Phys. Rev. B} {\bf 91}, 045425 (2015).

\bibitem{lahiri} J.~Lahiri, Y.~Lin, P.~Bozkurt, I.~I. Pleynik, and M.~Batzill,
\newblock {\em Nat. Nanotech.} {\bf 5}, 326 (2010).

\bibitem{td-ntwrk} J.~da~Silva-Ara\'ujo, H.~Chacham, and R.~W.~Nunes,
\newblock{\em Phys. Rev. B} {\bf 81}, 193405 (2010).

\bibitem{vozmediano} A.~Cortijo and M.~A.~H.~Vozmediano, \newblock{\em
Nucl. Phys. B} {\bf 763}, 293 (2007).

\bibitem{white} D.~Gunlycke and C.~T.~White, \newblock{\em
Phys. Rev. Lett.} {\bf 106}, 136806 (2011).

\bibitem{joice} J.~da Silva-Ara\'ujo and R.~W.~Nunes,
\newblock {\em Phys.~Rev.~B.} {\bf 81}, 073408 (2010).

\bibitem{louiegb1} O.~V.~Yazyev and S.~G.~Louie,
\newblock {\em Nat. Mater.} {\bf 9}, 806 (2010).

\bibitem{louiegb2} O.~V.~Yazyev and S.~G.~Louie,
\newblock {\em Phys. Rev. B} {\bf 81}, 195420 (2010).

\bibitem{tsen} A.~W.~Tsen, L.~Brown, M.~P.~Levendorf, F.~Ghahari,
P.~Y.~Huang, R.~W.~Havener, C.~S.~Ruiz-Vargas, D.~A.~Muller, P.~Kim,
and J.~Park, \newblock{\em Science} {\bf 336}, 1143 (2012).

\bibitem{yu} Q.~K.~Yu, L.~A.~Jauregui, W.~Wu, R.~Colby, J.~F.~Tian,
Z.~H.~Su, H.~L.~Cao, Z.~H.~Liu, D.~Pandey, D.~G.~Wei, T.~F.~Chung,
P.~Peng, N.~P.~Guisinger, E.~A.~Stach, J.~M.~Bao, S.~S.~Pei, and Y.~P.~Chen, 
{\em Nat. Mat.} {\bf 10}, 443 (2011).

\bibitem{simonis} P.~Simonis, C.~Goffaux, P.~A.~Thiry, L.~P.~Biro, 
P.~Lambin, and V.~Meunier, \newblock{\em Surf. Sci.} {\bf 511}, 319 (2002).

\bibitem{cervenka} J.~Cervenka and C.~F.~J.~Flipse, 
\newblock{\em Phys. Rev. B} {\bf 79}, 195429 (2009).

\bibitem{kim} K.~Kim, Z.~Lee, W.~Regan, C.~Kisielowski, M.~F.~Crommie, and A. Zettl,
\newblock{\em ACS Nano} {\bf 5}, 2142 (2011).

\bibitem{huang} P.~Y.~Huang, C.~S.~Ruiz-Vargas, A.~M.~van der Zande, W.~S.~Whitney,
M.~P.~Levendorf, J.~W.~Kevek, S.~Garg, J.~S.~Alden, C.~J.~Hustedt, Y.~Zhu, J.~Park,
P.~L.~McEuen, and D.~A.~Muller
\newblock{\em Nature} {\bf 469}, 389 (2011).

\bibitem{louie558} J.-H. Chen, G. Autès, N. Alem, F. Gargiulo,
A. Gautam, M. Linck, C. Kisielowski, O. V. Yazyev, S. G. Louie, and
A. Zettl.  \newblock {\em Phys. Rev. B} {\bf 89}, 121407(R) (2014).
  
\bibitem{stoner1} E.~C.~Stoner, \newblock{\em Proc. R. Soc. London Ser. A} {\bf 165}, 372 (1938).

\bibitem{stoner2} E.~C.~Stoner, \newblock{\em Proc. R. Soc. London Ser. A} {\bf 169}, 339 (1939).

\bibitem{clee} C.~Lee, X.~Wei, J.~W.~Kysar, and J.~Hone,
\newblock{\em Science} {\bf 321}, 385 (2008).

\bibitem{colombo} E.~Cadelano, P.~L.~Palla, S.~Giordano, and L.~Colombo,
\newblock{\em Phys. Rev. Lett.} {\bf 102}, 235502 (2009).

\bibitem{ahcn} V.~M.~Pereira and A.~H.~Castro Neto,
\newblock{\em Phys. Rev. Lett.} {\bf 103}, 046801 (2009).

\bibitem{paladio1} V.~L.~Moruzzi, and P.~M.~Marcus,
\newblock{\em Phys. Rev. B} {\bf 39}, 471 (1989).

\bibitem{paladio2} A.~Delin, E.~Tosatti, and R.~Weht,
\newblock{\em Phys. Rev. Lett.} {\bf 92}, 057201 (2004).

\bibitem{alexandre2} S.~S.~Alexandre, J.~M.~Soler, P.~J.~S.~Miguel,
R.~W.~Nunes, F.~Yndurain, J.~Gomez-Herrero, and F.~Zamora,
\newblock{\em Appl.~Phys.~Lett.} {\bf 90}, 193107 (2007).

\bibitem{alexandre3} S.~S.~Alexandre, M.~Mattesini, J.~M.~Soler, F.~Yndurain, 
\newblock{\em Phys. Rev. Lett.} {\bf 96}, 079701 (2006).

\bibitem{alexandre4} S.~S.~Alexandre, E.~Anglada, J.~M.~Soler, F.~Yndurain, 
\newblock{\em Phys. Rev. B} {\bf 74}, 54405 (2006).

\bibitem{mermim-wagner} N.~D.~Mermin and H.~Wagner,
\newblock{\em Phys. Rev. Lett.} {\bf 17}, 1133 (1996).

\bibitem{ks} W.~Kohn and L.~J.~Sham,
\newblock{\em Phys.~Rev.} {\bf 140}, A1133 (1965).

\bibitem{gga} J.~P.~Perdew, K.~Burke, and M.~Ernzerhof,
\newblock{\em Phys. Rev. Lett.} {\bf 77}, 3865 (1996).

\bibitem{tm-kb1} N.~Troullier and J~.L.~Martins,
\newblock{\em Phys.~Rev.~B} {\bf 43}, 1993 (1991).

\bibitem{tm-kb2} L.~Kleinman and D.~M.~Bylander,
\newblock{\em Phys.~Rev.~Lett.} {\bf 48}, 1425 (1982).

\bibitem{siesta} J.~M.~Soler, E.~Artacho, J.~D.~Gale, A.~Garcia,
J.~Junquera, P.~Ordejon, and D.~S\'anchez-Portal,
\newblock{\em J.~Phys.~Cond.~Matt.} {\bf 14}, 2745 (2002).

\end{thebibliography}
\end{document}